\documentclass[reprint,showpacs,amsmath,amssymb,aps,prb,superscriptaddress]{revtex4-1}

\usepackage{graphicx}
\usepackage{dcolumn}
\usepackage{bm}

\begin{document}

\title{First-principles study on the electronic and transport properties of periodically nitrogen-doped graphene and carbon nanotube superlattices}

\author{Fuming Xu}
\affiliation{College of Physics and Energy, Shenzhen University, Shenzhen 518060, China}
\author{Zhizhou Yu}\email{zzyu@connect.hku.hk}
\affiliation{School of Physics and Technology, Nanjing Normal University, Nanjing 210023, China}
\affiliation{Department of Physics and the Center of Theoretical and Computational Physics, The University of Hong Kong, Hong Kong, China}
\author{Zhirui Gong}
\affiliation{College of Physics and Energy, Shenzhen University, Shenzhen 518060, China}
\author{Hao Jin}
\affiliation{College of Physics and Energy, Shenzhen University, Shenzhen 518060, China}

\begin{abstract}
Prompted by recent reports on $\sqrt{3} \times \sqrt{3}$ graphene superlattices with intrinsic inter-valley interactions, we perform first-principles calculations to investigate the electronic properties of periodically nitrogen-doped graphene and carbon nanotube nanostructures. In these structures, nitrogen atoms substitute one-sixth of the carbon atoms in the pristine hexagonal lattices with exact periodicity to form perfect $\sqrt{3} \times \sqrt{3}$ superlattices of graphene and carbon nanotubes. Multiple nanostructures of $\sqrt{3} \times \sqrt{3}$ graphene ribbons and carbon nanotubes are explored, and all configurations show nonmagnetic and metallic behaviors. The transport properties of $\sqrt{3} \times \sqrt{3}$ graphene and carbon nanotube superlattices are calculated utilizing the non-equilibrium Green's function formalism combined with density functional theory. The transmission spectrum through the pristine and $\sqrt{3} \times \sqrt{3}$ armchair carbon nanotube heterostructure shows quantized behavior under certain circumstances.
\end{abstract}


\maketitle

\section{Introduction}\label{sec1}

As the first artificially fabricated two-dimensional material, graphene exhibits extraordinary properties and is potentially suitable for a wide range of applications \cite{graphene0,graphene1,graphene2,graphene3,graphene4,graphene5,graphene6,graphenefp,graphenefp2}. The honeycomb lattice provides the Bloch electrons in graphene a new degree of freedom, \textit{valley}, which can be manipulated to store and process binary information\cite{valley05,valley06,valley071,valley072}. Therefore, graphene has been extensively studied as a promising valleytronic material. Multiple setups have been proposed that utilize the intrinsic characteristics of graphene nanostructures, such as zigzag edges\cite{valleygraphene1}, 5-7-5 line defects\cite{valleygraphene21,valleygraphene22,valleygraphene23,valleyjiang}, zero-line modes\cite{valleyzline0,valleyzline1,valleyzline2}, or via extrinsic methods, such as strain engineering\cite{valleygraphene311,valleygraphene312,valleygraphene321,valleygraphene322} and temperature-gradient driving\cite{valleygraphene41,valleygraphene42}, to tune the valley-related currents through graphene-based nanostructures.

Recently, it was found that graphene superlattices with certain periodicities possess intrinsic inter-valley interactions. For instance, theoretical work revealed that in $\sqrt{3}N \times \sqrt{3}N$ and $3N \times 3N$ graphene superlattices, the band folding merges the valley degree of freedom in pristine graphene, and universal inter-valley couplings naturally arise\cite{qiao15}. Similar to the role of spin-orbit coupling in spintronics, the inter-valley interaction can serve as valley-orbit couplings to effectively process the valley information, which qualifies these graphene superlattices as prospective valley-processing units in future integrated valleytronic circuits. It was suggested that these graphene superlattices could be realized through top-absorption or periodical doping on pristine graphene. Moreover, the latest investigations found that, owing to the proximity effect, graphene on top of a topological-insulator substrate naturally shows these types of superlattice patterns\cite{grapheneTI13,grapheneTI14}. Based on the effective Hamiltonian proposed in Ref.~[\onlinecite{qiao15}], the electronic and valleytronic properties of multiple graphene and carbon nanotube (CNT) nanostructures with $\sqrt{3} \times \sqrt{3}$ superlattice were discussed on tight-binding level\cite{xu16}. A valley-field-effect-transistor containing $\sqrt{3} \times \sqrt{3}$ armchair CNTs with outstanding device functionality was presented in Ref.~[\onlinecite{xu16}].

Nitrogen doping has been widely used in laboratory settings in an attempt to improve the carrier density and conductivity of graphene systems\cite{nitrogen1,nitrogen2,nitrogen3}. Selective adsorption of ammonia molecules at the edges of zigzag graphene nanoribbons has been achieved\cite{nitrogen4}, which offers opportunities for the precise control of dopant positions in graphene systems. Hypothetically, if the dopant nitrogen atoms can be precisely controlled to periodically locate in the host graphene lattice, a favorable graphene superlattice with intrinsic inter-valley interactions is constructed. Considering the substitution case, and taking the representative $\sqrt{3} \times \sqrt{3}$ superlattice as an example, nitrogen atoms will substitute  one-sixth of the carbon atoms at the same positions in each honeycomb ring, as illustrated in Fig.{\ref{fig1}}. At present, it is difficult to realize this structure; however, first-principles calculations can elucidate the properties of the material in advance.

In this article, we present first-principles investigations on the electronic properties of typical $\sqrt{3} \times \sqrt{3}$ graphene and single wall CNT superlattices with periodic nitrogen dopants. The electronic band structures of multiple graphene ribbons and single wall CNT superlattices with typical zigzag and armchair chiralities are studied, including three types of zigzag ribbons, two types of armchair ribbons, as well as zigzag and armchair tubes. All these superlattice nanostructures exhibit nonmagnetic and metallic characteristics. The nitrogen-doped graphene ribbons have intrinsic band gaps of finite-size nature, and a flat band resides in the bulk gap in two types of zigzag ribbons. The zigzag tube superlattice has a tiny energy gap at small system sizes, while no band gap exists in armchair nanotubes. The transport properties of heterostructures consisting of pristine and $\sqrt{3} \times \sqrt{3}$ graphene ribbons and CNTs are explored, adopting the non-equilibrium Green's function formalism combined with density functional theory. The transmission spectra are spin dependent only near the transmission gap in the case of zigzag graphene ribbons. Moreover, a quantized transmission plateau occurs in the transmission spectrum of a metallic armchair CNT heterostructure.

The rest of the paper is organized as follows. In Section {\ref{sec2}}, we introduce the computational details as well as the geometrical structures of multiple nitrogen-doped graphene nanoribbons and CNTs with $\sqrt{3} \times \sqrt{3}$ superlattice, which will be investigated in the electronic and transport calculations. In Section {\ref{sec3}}, we present the electronic structures of these graphene nanoribbons and CNTs superlattices, and their transport properties are outlined. Finally, a brief summary of the main results is provided in Section {\ref{sec4}}. For simplicity, we refer to the periodically nitrogen-doped graphene with $\sqrt{3} \times \sqrt{3}$ superlattice as {\it nitrogen-doped graphene} or {\it $\sqrt{3} \times \sqrt{3}$ graphene superlattice} in the following text. Similar abbreviations also apply for the CNT case.

\section{Computational Methods and Geometric Models}\label{sec2}

To investigate the structural and electronic properties of $\sqrt{3} \times \sqrt{3}$ graphene and CNT superlattices, first-principles calculations are performed on a plane-wave basis with the projector augmented wave (PAW)\cite{PAW} to model the electron-ionic core interaction, as implemented in the Vienna Ab initio Simulation Package (VASP)\cite{vasp1,vasp2}. The exchange and correlation interactions are approximated by the generalized gradient approximation (GGA) with the Perdew-Burke-Ernzerhof (PBE) functional\cite{PBE}. A plane-wave basis set with a kinetic energy cutoff of 400 eV is employed. In the calculations, 10 Monkhost-Pack \textit{k}-points are used along the 1-dimensional Brillouin zone for the zigzag graphene ribbons and armchair CNTs, while 8 \textit{k}-points are sampled for the armchair graphene ribbons and zigzag CNTs. All atoms in the supercell are fully relaxed with a residual force less than 0.02 eV/{\AA}, and the total energies are converged to $10^{-5}$ eV. The transport properties of the two-probe systems are investigated using NanoDcal, which is a first-principles package within the non-equilibrium Green's function formalism\cite{nanodcal}. The double $\zeta$ plus polarization numerical orbital basis set is used. The mesh cutoff is set to be 3,000 eV and the convergence of the total energies is $10^{-5}$ eV.

\begin{figure}[tbp]
\includegraphics[width=\columnwidth]{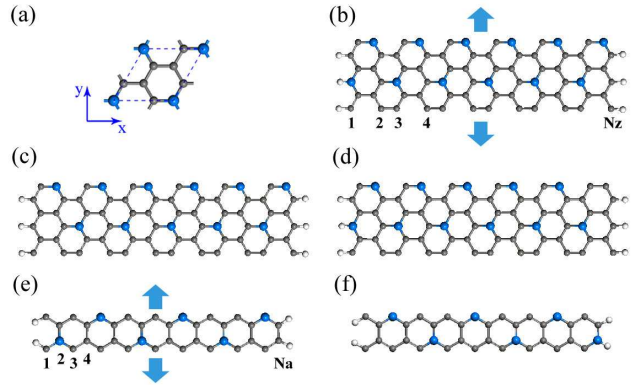}
\caption{Top views of the geometric structures of (a) 2-dimensional nitrogen-doped graphene superlattice, (b) 12-zigzag ribbon (ZGR1), (c) 12-zigzag ribbon (ZGR2), (d) 12-zigzag ribbon (ZGR3), (e) 18-armchair ribbon (AGR1), and (f) 18-armchair ribbon (AGR2). The nanoribbons are extended periodically along the $y$ direction, as indicated by the blue arrows. The white, gray, and blue balls represent the hydrogen, carbon, and nitrogen atoms, respectively.}
\label{fig1}
\end{figure}

Figure ~\ref{fig1}(a) shows the primitive cell of the bulk $\sqrt{3} \times \sqrt{3}$ graphene superlattice, which contains five carbon atoms and one nitrogen atom. The optimized in-plane lattice constant in our calculation is 4.22 \AA. The models of graphene ribbons are constructed by cutting the 2-dimensional bulk graphene superlattice with different edges and widths. All nanoribbons are hydrogen terminated in the calculations. According to the previous studies of pristine graphene ribbons\cite{Louie1}, the widths of zigzag and armchair nitrogen-doped graphene ribbons are denoted by the numbers of zigzag chains (N$_\textrm{z}$ in Fig.~\ref{fig1}(b)) and dimmer lines (N$_\textrm{a}$ in Fig.~\ref{fig1}(e)) across the ribbon width, respectively. For instance, N$_\textrm{z}=12$ in Fig.~\ref{fig1}(b) and N$_\textrm{a}=18$ in Fig.~\ref{fig1}(e).

Figure~\ref{fig1}(b) presents the geometric structures of the 12-zigzag ribbon cut directly from the graphene superlattice, which consists of six dimmer lines along the periodical direction in one supercell. Nitrogen atoms locate on both zigzag edges in this configuration, which we refer to as ZGR1. The calculation of total energy suggests that the geometry of ZGR1 was less stable owing to the presence of the nitrogen atoms on both edges. Therefore, we constructe two different configurations of zigzag graphene ribbons by substituting nitrogen atoms with carbon atoms along one edge of ZGR1 to form one pure carbon edge. We refer to these two structures as ZGR2 and ZGR3, as shown in Figs.~\ref{fig1}(c) and \ref{fig1}(d), respectively. The molar Gibbs free energy of formation\cite{Barone,note} of ZGR3 from our first-principles calculation is 11 meV/atom less than that of ZGR1, and 8 meV/atom lower than that of ZGR2. This fact proves that ZGR3 is the most stable configuration of zigzag nitrogen-doped graphene ribbons. Similarly, two different configurations of armchair ribbons of graphene superlattice, referred to as AGR1 and AGR2, are presented in Figs.~\ref{fig1}(e) and \ref{fig1}(f), respectively. AGR2 is more stable since its total energy is 6 meV/atom less than that of AGR1. When wrapping AGR1 or AGR2 with the same ribbon width, two identical zigzag CNT superlattices are obtained. Meanwhile, a $\sqrt{3} \times \sqrt{3}$ armchair CNT is realized by rolling up ZGR1. These CNT superlattices have the same chemical composition as the bulk nitrogen-doped graphene except for the terminal hydrogen atoms, and we denote them by a pair of indices $(n,m)$ as those used for pristine CNTs\cite{CNTreview1,CNTreview2}. The spin degree of freedom is neglected in the electronic structure calculations since these hydrogen-saturated $\sqrt{3} \times \sqrt{3}$ graphene ribbons and CNTs are found to have nonmagnetic ground states in total energy computations.

\begin{figure}[tbp]
\includegraphics[width=\columnwidth]{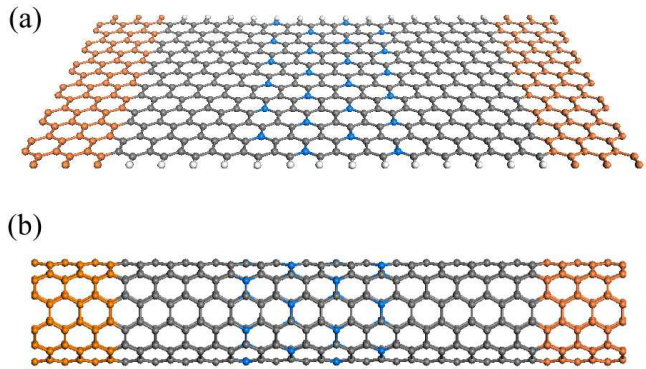}
\caption{Two-probe configurations of (a) 12-zigzag nitrogen-doped graphene ribbon, and (b) (6,6)-armchair nitrogen-doped CNT sandwiched between pristine zigzag graphene ribbons and armchair CNTs, respectively. The white, gray, and blue balls denote the hydrogen, carbon, and nitrogen atoms, respectively. The orange regions represent the lead regions.}
\label{fig2}
\end{figure}

These $\sqrt{3} \times \sqrt{3}$ nitrogen-doped graphene and CNT superlattices may serve as valley-processing units in valleytronic applications. Therefore, to study the transport characteristics of the $\sqrt{3} \times \sqrt{3}$ graphene nanoribbons and CNTs, we construct two-probe structures by sandwiching these superlattices between electrodes made of either pristine graphene ribbons or CNTs with the same system sizes. Panels (a) and (b) of Fig.~\ref{fig2} depict the two-probe configurations of 12-$\sqrt{3} \times \sqrt{3}$ zigzag graphene ribbon with pristine graphene ribbon leads, and (6,6)-$\sqrt{3} \times \sqrt{3}$ armchair CNTs with pristine CNT leads, respectively. As clearly demonstrated in this figure, the ribbon system is hydrogen-saturated while the CNT setup has cylindrical boundary conditions. The structures are also fully relaxed in VASP. Since pristine zigzag graphene ribbons exhibit antiferromagnetic ground states\cite{Louie1}, spin-resolved transport properties are evaluated in the transport calculations of graphene ribbon two-probe systems.

\section{Numerical Results and Discussions}\label{sec3}

We carry out extensive first-principles calculations on the electronic and transport properties of carbon nanostructures containing $\sqrt{3} \times \sqrt{3}$ superlattices, and the numerical results are presented in detail.

\subsection{Electronic Properties of $\sqrt{3} \times \sqrt{3}$ nitrogen-doped graphene nanoribbons and CNTs}

\begin{figure}[tbp]
\includegraphics[width=\columnwidth]{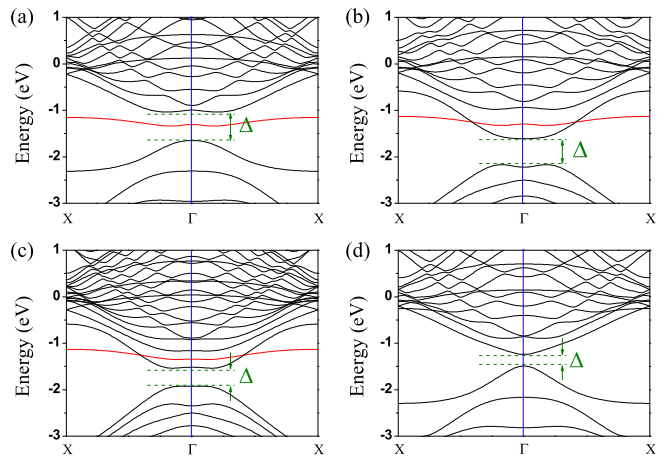}
\caption{Electronic band structures of $\sqrt{3} \times \sqrt{3}$ nitrogen-doped graphene ribbons: (a) 12-ZGR1, (b) 12-ZGR2, (c) 18-ZGR2, and (d) 12-ZGR3. The corresponding geometries are shown in Fig.~\ref{fig1}.}
\label{fig3}
\end{figure}

First, the band structures of nitrogen-doped zigzag graphene ribbons of different configurations are illustrated in Fig.~\ref{fig3}. Because of the band folding of the $\sqrt{3} \times \sqrt{3}$ superlattice, all bands are centered at the $\Gamma$ point instead of $K/K'$ in pristine graphene. Compared with the Dirac cone bands at the $K/K'$ points of pristine graphene, electrons in these $\sqrt{3} \times \sqrt{3}$ zigzag ribbons travel with lower group velocities near the Fermi energy. All three zigzag ribbons are metallic, with intrinsic energy gaps below the Fermi energy locate at $E\approx-1.5$ eV. This behavior is straightforward since nitrogen is widely used as an \textit{n}-type dopant to shift the Fermi level up. We label the magnitude of this intrinsic energy gap as $\Delta$. For the 12-ZGR1, an indirect energy gap of $\Delta=0.61$ eV emerges with its top and bottom positions residing at the $\Gamma$ point and approximately 1/5th of the $\Gamma$-$X$, as shown in Fig.~\ref{fig3}(a). A nearly flat band, highlighted by the red line, exists at the center of the energy gap and primarily originates from the edge carbon atoms at the all-carbon side (right-hand side of Fig.~\ref{fig1}(b)). This flat band is similar to the edge states of pristine zigzag graphene ribbons. In Fig.~\ref{fig3}(b), the indirect gap of 12-ZGR2 is 0.56 eV between $\Gamma$-$X$ at the bottom and $\Gamma$ point at the top. The bulk energy gap is approximately half an electron volt lower than that of 12-ZGR1. A flat band exists in the bulk conduction bands, which is mostly caused by the edge state of the right-side zigzag carbon chain with nitrogen, as shown in Fig.~\ref{fig1}(c). When the ribbon width of ZGR2 increases, the bulk gap decreases. But the flat band is almost independent of the system size, as shown in Fig.~\ref{fig3}(c). However, when N$_\textrm{z}$ exceeds 16, as the case for 18-ZGR2, even though the energy gap of 0.39 eV is still indirect, the top and bottom positions move to $\Gamma$ point and close to $\Gamma$ at $\Gamma$-$X$, respectively. In other words, ZGR2 gradually evolves from an indirect to a direct gap material when the ribbon width grows. In contrast to ZGR1 and ZGR2, ZGR3 has a smaller direct band gap at the same size. Figure~\ref{fig3}(d) shows $\Delta=0.25$ eV at $\Gamma$ point in 12-ZGR3. The nearly flat band is absent in ZGR3 because the corresponding edge state is broken. With the exception of these differences, all nitrogen-doped zigzag graphene ribbons share common features owing to their close geometries. Most of these first-principles results are qualitatively consistent with the tight-binding predictions in Ref.[\onlinecite{xu16}].
\begin{figure}
\centering
\includegraphics[width=\columnwidth]{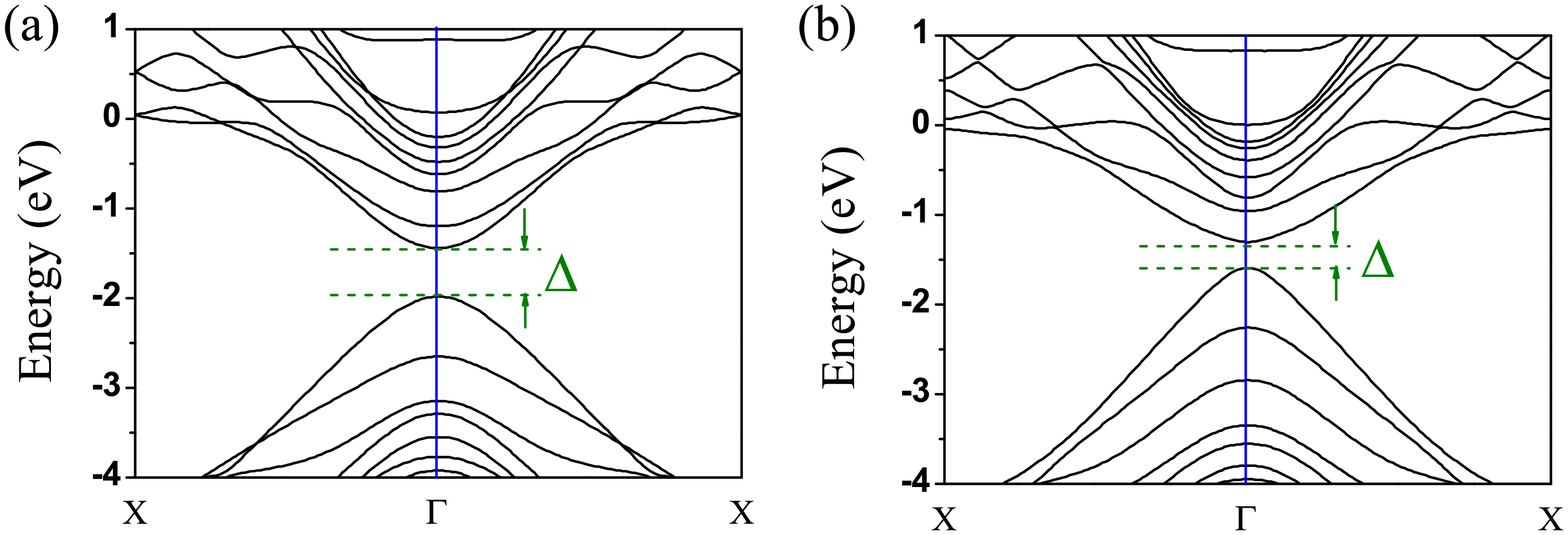}
\caption{Electronic band structures of $\sqrt{3} \times \sqrt{3}$ nitrogen-doped armchair graphene ribbons, (a) 18-AGR1, and (b) 18-AGR2. The spatial configurations of AGR1 and AGR2 are visualized in Fig.~\ref{fig1}(e) and Fig.~\ref{fig1}(f), respectively.}
\label{fig4}
\end{figure}

Second, numerical results on the dispersion relations of $\sqrt{3} \times \sqrt{3}$ armchair graphene ribbons are shown in Fig.~\ref{fig4}, which resemble the metallic features of nitrogen-doped zigzag graphene ribbons. Both AGR1 and AGR2 have direct band gaps at $\Gamma$ point where all of the electrical bands focus. The intrinsic energy gaps of 18-AGR1 and 18-AGR2 are 0.55 eV and 0.29 eV, respectively. These bulk gaps also reside around $E\approx -1.5$ eV in the $E-k$ maps. Similar to the pristine armchair graphene ribbons case, there is no edge mode or flat band in the dispersion relations of these $\sqrt{3} \times \sqrt{3}$ armchair graphene ribbons. The band gap of AGR2 is drastically smaller than that of AGR1 at the same ribbon width, and AGR2 is also more structurally stable than AGR1.

\begin{figure}[tbp]
\centering
\includegraphics[width=\columnwidth]{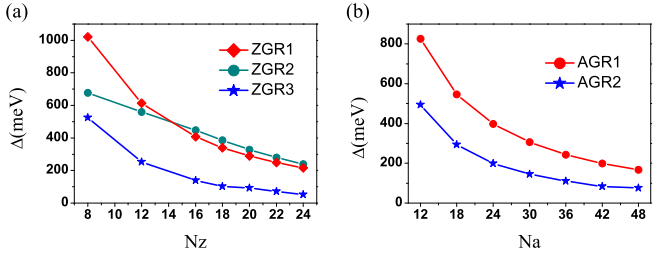}
\caption{(a) Evolution of band gap $\Delta$ of the nitrogen-doped zigzag graphene ribbons as a function of ribbon width N$_\textrm{z}$. (b) Band gap $\Delta$ of $\sqrt{3} \times \sqrt{3}$ armchair ribbons versus ribbon width N$_\textrm{a}$.}
\label{fig5}
\end{figure}

Moreover, relations between the ribbon width and bulk gap $\Delta$ are studied, and the tendencies are visualized in Fig.~\ref{fig5}. We calculate the band structures of a series of zigzag ribbons with increasing widths, and plot the corresponding $\Delta$ as a function of ribbon width in Fig.~\ref{fig5}(a). For a narrow ribbon ($N_z < 14$), ZGR1 has the largest gap, while $\Delta$ of ZGR3 is the smallest. When increasing the ribbon width, the band gap $\Delta$ decreases for all three types of zigzag ribbons. The band gaps of both ZGR1 and ZGR3 exponentially decay when $N_z$ grows, but $\Delta$ of ZGR2 shows a linear decreasing relation with $N_z$. As a result, ZGR2 has the largest band gap for $N_z > 14$. The decaying behavior of the band gaps indicates their finite-size nature, which will eventually vanish at large system sizes. We notice that ZGR3 not only has the smallest gap, but also is the most stable structure among these zigzag ribbons. We have also studied the energy gap of $\sqrt{3} \times \sqrt{3}$ armchair graphene ribbons at various ribbon widths, as presented in Fig.~\ref{fig5}(b). The bulk gaps of both AGR1 and AGR2 exponentially decay as the ribbon width N$_\textrm{a}$ increases. Analogous to the case of zigzag ribbons, $\Delta$ of the more stable configuration AGR2 is always smaller than that of AGR1. The finite-size gap of the armchair ribbons is shown to be rather robust against the system size. For instance, $\Delta$ is around $100$ meV at $N_a=48$ for both AGR1 and AGR2 systems.

\begin{figure}[tbp]
\centering
\includegraphics[width=\columnwidth]{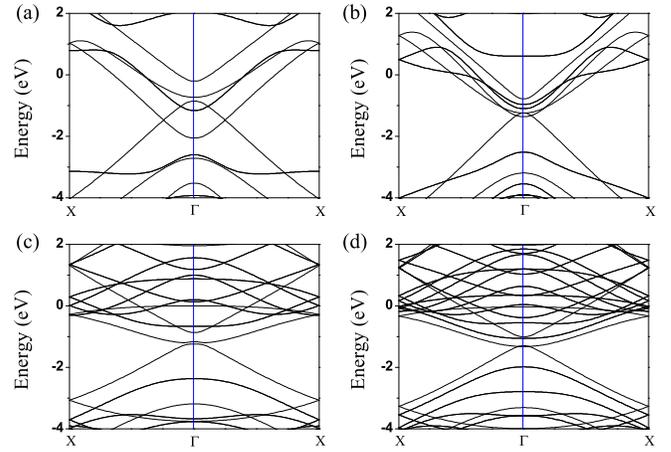}
\caption{Electronic band structures of $\sqrt{3} \times \sqrt{3}$ nitrogen-doped single wall CNTs: (a) (6,0) zigzag CNT, (b) (9,0) zigzag CNT, (c) (6,6) armchair CNT, and (d) (9,9) armchair CNT.}
\label{fig6}
\end{figure}

Finally, we explore the electronic band structures of nitrogen-doped single wall CNTs with $\sqrt{3} \times \sqrt{3}$ superlattices. Typical chirality of zigzag and armchair is investigated, and numerical results are shown in Fig.~\ref{fig6}. It is clear that all of the nitrogen-doped CNTs considered here exhibit metallic characteristics. Unlike nitrogen-doped graphene ribbon superlattices, these CNT superlattices do not have significant band gaps in their $E-k$ relations. As for zigzag CNT, no band gap can be observed in Figs.~\ref{fig6}(a) and \ref{fig6}(b). For a smaller $(6,0)$ zigzag CNT, remarkable band crossings occur symmetrically around the $\Gamma$ point at $E \approx -1.47$ eV in Fig.~\ref{fig6}(a). When the tube size increases to $(9,0)$, the band crossings gradually evolve into band touches at the $\Gamma$ point, as shown in Fig.~\ref{fig6}(b). Another distinct feature in the band structures of these zigzag CNTs is the linear bands, which guarantee high group velocities of the propagating Bloch electrons. Linear bands can lead to the counterintuitive Klein tunneling phenomenon, where high-speed incoming electrons can normally penetrate a potential barrier with perfect transmission. The existence of Klein tunneling behavior has been confirmed in pristine CNTs\cite{KleinCNT1,KleinCNT2,KleinCNT3}, and Ref.[\onlinecite{xu16}] predicted its existence in $\sqrt{3} \times \sqrt{3}$ CNT superlattices on tight-binding level. As for armchair tubes, a tiny energy gap exists in the case of small tubes. For example, Fig.~\ref{fig6}(c) demonstrates that a $(6,6)$ armchair CNT has an indirect gap of $45$ meV near the $\Gamma$ point. But this gap quickly closes as the tube size grew to $(9,9)$, and band touching can be seen in Fig.~\ref{fig6}(d). We also find several linear bands embedded in the bulk band structures. Compared with those of zigzag CNT superlattices, the bands of armchair tubes are much denser at similar tube diameters.

\subsection{Transport Properties of Nitrogen-doped Graphene and CNT Heterostructures}

We investigate the transport properties of nitrogen-doped graphene and CNT superlattices using the NanoDcal package, which is currently the most popular first-principles quantum transport simulation software. In particular, we calculated the transmission coefficients through two-probe heterostructures consisting of nitrogen-doped graphene ribbon (nitrogen-doped CNT) sandwiched between pristine graphene ribbon (pristine CNT) leads, which are shown in Fig.~\ref{fig2}.

\begin{figure}[tbp]
\centering
\includegraphics[width=\columnwidth]{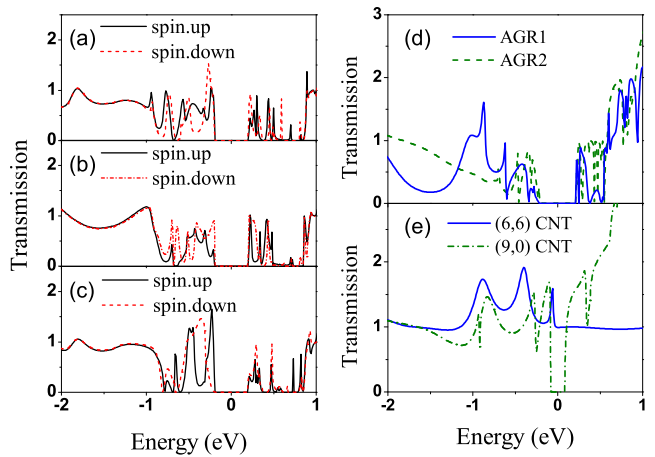}
\caption{Transmission versus energy through pristine and nitrogen-doped two-probe systems, (a) 12-ZGR1, (b) 12-ZGR2, (c) 12-ZGR3, (d) 18-AGR1, and 18-AGR2, and (e) (6,6) and (9,0) CNT superlattices, respectively.}
\label{fig7}
\end{figure}

The zigzag graphene ribbon systems are studied, and numerical results are provided in Fig.~\ref{fig7}. We choose a 12-zigzag ribbon and an 18-armchair ribbon as typical configurations, and then construct two-probe systems. The pristine zigzag graphene ribbon has a small band gap within its antiferromagnetic ground state\cite{Louie1}. When pristine zigzag graphene ribbons are used as electrodes, a zero-transmission gap of $0.38$ eV appears in the spin-resolved transmission coefficients of all two-probe configurations with a system width of $N_z = 12$. The corresponding diagrams for nitrogen-doped ZGR1, ZGR2, and ZGR3 two-probe systems are displayed in panels (a), (b), and (c) of Fig.~\ref{fig7}, respectively. Except for the same zero-transmission gap, another common feature among these figures is that the transmission coefficients are only spin-dependent in the energy range of $E \in (-1, 1)$ eV, which corresponds to the $\pi$ and $\pi^*$ channels for all pristine zigzag graphene based systems\cite{graphene2008}. We also observe many spin-dependent resonant peaks in this range. As for the 12-ZGR1 system, the transmission peak of spin-down electrons from the highest occupied molecular orbital (HOMO) is larger than that of spin-up electrons, and exceeds 1. Both peaks are suppressed in the ZGR2 system, in which the transmission of spin-down electrons is still larger. In contrast, for the 12-ZGR3 system, the transmission peak of spin-down electrons moves to a lower energy, and the transmission at the HOMO is spin-up polarized. Both of these transmission peaks are greater than one in magnitude.

Figure.~\ref{fig7}(d) shows the transmission coefficients of two-probe armchair graphene ribbon systems with width $N_a=18$. Both of the pristine and nitrogen-doped armchair graphene ribbons have nonmagnetic ground states. Therefore, spin is not involved in the calculation. Owing to the semiconducting nature of pristine armchair graphene ribbons, the 18-AGR1 and 18-AGR2 systems possess transmission gaps of 0.42 eV around $E=0$. In the $E > 0$ region, almost quantized transmission peaks appears in both systems. In the case of $E < 0$, a step-increasing behavior of transmission is observed in the AGR1 two-probe system, while transmission peaks appear at the HOMO for the AGR2 system. Below $E=-1$ eV, both systems have smooth transmission curves. It is worth noting that the transmission function of the 18-AGR2 system shows strong oscillations, and several transmission peaks approach one near $E \approx 0.4 eV$.

The transmissions of CNT based two-probe systems are also evaluated. We consider the representative armchair and zigzag chiralities and present the numerical results in Fig.~\ref{fig7}(e). Spin polarizations are absent in these CNT systems. Typical $(6,6)$ and $(9,0)$ tubes are employed to build two-probe systems, with pristine CNTs as electrodes and nitrogen-doped CNT superlattices as the scattering region. The $(6,6)$ armchair CNT system exhibits conductive behavior across the entire energy range. Remarkable quantized transmission plateaus of $T = 1$ are found in the energy ranges $E \in [0, 1]$ eV and $E \in [-1.75, -1.15]$ eV, which are the symbol of perfect ballistic transport. Between the two plateaus, the transmission function slowly fluctuates from one to two in the region $E \in [-1, 0]$ eV. In the case of a $(9,0)$ zigzag CNT system, there is a transmission gap of 0.16 eV, which originates from the semiconducting electrodes. The transmission curve has a smooth shape below $E = -1$ eV, and wildly oscillates from zero to more than three in the region $E \in [-1, 1]$ eV. The relatively large transmission of these tube systems may be a direct result of the linear bands shown in Fig.\ref{fig6}.

\section{Summary}\label{sec4}

In summary, we have performed first-principles investigations on the electronic and transport properties of nitrogen-doped zigzag and armchair graphene ribbons and carbon nanotubes with $\sqrt{3} \times \sqrt{3}$ superlattices, which have been reported to possess intrinsic inter-valley interactions. The electronic band structures reveal that all nitrogen-doped graphene ribbons and carbon nanotubes are nonmagnetic metals. Three types of zigzag ribbons and two kinds of armchair ribbons are studied and they all have finite-size band gaps below the Fermi energy, which gradually close as the ribbon width increases. One nearly flat band originated from the edge carbon atoms resides inside the bulk gap for two configurations of zigzag ribbons. Numerical results suggest that stable ribbons with lower Gibbs free energies also exhibit smaller band gaps in both zigzag and armchair cases. In contrast, nitrogen-doped carbon nanotube superlattices have no band gap in the case of zigzag chirality, and a tiny gap of dozens meV exists in small armchair tubes, which quickly vanishes as the tube size grows. Several linear bands are identified in the dispersion relations of nitrogen-doped carbon nanotubes, which implies that these tubes have extraordinary ballistic transport properties. The transport properties of pristine and nitrogen-doped graphene ribbon/carbon nanotube heterostructures are investigated using the first-principles method. In particular, we have calculated the transmission coefficients as functions of the energy through two-probe systems. Two-probe systems constructed on zigzag ribbons, armchair ribbons, and zigzag tubes exhibit semiconducting characteristics with transmission gaps. Spin-resolved resonant peaks are observed around the transmission gaps in the zigzag ribbon systems. The armchair carbon nanotube heterostructures show metallic behavior, and remarkable quantized transmission plateaus of $T=1$ appear in wide energy ranges in the (6,6) CNT system. Given these extraordinary properties, the $\sqrt{3} \times \sqrt{3}$ zigzag and armchair graphene ribbons and carbon nanotubes could be promising materials for future valleytronic applications.

\begin{acknowledgments}
This work was financially supported by the National Natural Science Foundation of China (Grants No. 11504240, 11504241, and 11604213), and Natural Science Foundation of Shenzhen University (Grant No. 201550). Z.~Y. acknowledges the University Grant Council (Contract No. AoE/P-04/08) of the Government of HKSAR.
\end{acknowledgments}


\begin{thebibliography}{34}
\bibitem{note}
The chemical potentials of hydrogen, nitrogen and carbon are chosen as the binding energy per atom of the H$_2$ and N$_2$ molecules, and the cohesive energy per atom of a single graphene sheet, respectively.
\end{thebibliography}
\end{document}